\documentclass[twocolumn]{aastex631}
\usepackage{float}

\graphicspath{{./}{figures/}}

\begin{document}

\title{Radio emission from the magnetically active M dwarf UV Ceti from 1 GHz to 105 GHz}  

\author{Kathryn Plant}
\affiliation{Division of Physics, Mathematics, and Astronomy, California Institute of Technology, 
Pasadena, California 91125, USA}
\author{Gregg Hallinan}
\affiliation{Division of Physics, Mathematics, and Astronomy, California Institute of Technology, 
Pasadena, California 91125, USA}
\author{Tim Bastian}
\affiliation{National Radio Astronomy Observatory, 520 Edgemont Road, Charlottesville, VA 22903 USA}

\begin{abstract}

BL and UV Ceti are a nearby (2.7 pc) binary system with similar masses, spectral types, and rapid rotation rates, but very different magnetic activity. UV Ceti's much stronger large-scale magnetic field may cause this difference, highlighting key unanswered questions about dynamo processes in fully convective objects. Here we present multi-epoch characterization of the radio spectrum of UV Ceti spanning 1--105 GHz, exhibiting flared emission similar to coronal activity, auroral-like emission analogous to planetary magnetospheres, and slowly-varying persistent emission. Radio observations are a powerful means to probe the role that the large-scale magnetic field of UV Ceti has in non-thermal particle acceleration, because radio-frequency phenomena result from both the activity of small-scale field features as well as large-scale auroral current systems. We find temporal variability at all bands observed, and a hint of rotational modulation in the degree of circular polarization up to 40\,GHz. The persistent component of the emission is fairly constant from 1--105\,GHz, making optically thick emission or optically thin gyrosynchrotron from electrons with an isotropic pitch angle distribution unlikely.  We discuss the possibility of emission mechanisms analogous to Jupiter's radiation belts.

\end{abstract}

\section{Introduction}

M-dwarfs cooler than M3 or M4 are expected to be fully convective and thus must generate their magnetic fields with a different type of dynamo than the Sun \citep{chabrier1997fullyconvective}, whose dynamo depends on the tachocline boundary layer between a convective outer region and a radiative core \citep[e.g.][]{parker1955dynamo}.
Many M dwarfs exhibit far more intense magnetic activity than the present-day Sun, as evinced by:  more frequent and more powerful flares \citep[e.g.][]{Feinstein2020flare-rates}, kiloGauss magnetic fields with large filling factors \citep{Johns-Krull1996ZDImdwarfkGfields,Morin2008,Donati2008}, higher ratio of X-ray and H$\alpha$ luminosity to bolometric luminosity \citep[e.g.][]{Wright2011activity-rotation}, and persistent nonthermal radio emission \citep[e.g.][]{Linsky1983}.

This magnetic activity persists for much longer than it does in the life-cycle of a sun-like star \citep{West2008magnetic_activity_lifetimes}, consistent with slower spin-down timescales for fully-convective M dwarfs.
Faster-rotating stars tend to be more magnetically active than slowly-spinning stars. This relation results in a monotonic increase in the ratio of X-ray luminosity to bolometric luminosity  ($\frac{L_x}{L_B}$) with an increase in the stellar Rossby number (ratio of rotation period to convective turnover timescale), a trend which saturates at the shortest rotation periods with $Log(L_x/L_B)\sim -3$  (for a summary see \cite{Wright2011activity-rotation} and references therein).

One of the observable key differences between the dynamos in sun-like stars and fully convective M dwarfs is the prevalence of strong (up to a few kiloGauss) large-scale magnetic fields in M dwarfs, as revealed by Zeeman Doppler Imaging
\citep[ZDI;][]{Donati2008,Morin2008}. It is becoming increasingly apparent that such large-scale magnetic fields can underlie differences in the magnetic activity observed.
Large-scale magnetospheres can sustain stable current-systems that transport energy from the middle and outer magnetosphere into the lower atmosphere, resulting in auroral emissions (both radio and optical) from M dwarfs at the end of the main sequence as well as brown dwarfs (collectively termed ultracool dwarfs) \citep{hallinan2015aurora,Hallinan2006,Kao2018}. The magnetic field configuration may also explain the bimodal distribution in spindown rates of M dwarfs \citep{Garraffo2018spindown}.

The radio signature of auroral activity in ultracool dwarfs is the presence of periodic, highly circularly polarized radio emission, believed to be electron cyclotron maser emission (ECME) produced at either the local gyrofrequency, or the first harmonic \citep[e.g.][]{hallinan2015aurora, Kao2018, Villadsen2019, Zic2019, Lynch2017ECME}. It is detected from $\sim 20\%$ ultracool dwarfs ranging from spectral type M8 to T8. 
Although large optical flares are still observed from some ultracool dwarfs \citep[e.g.][]{Paudel2020Ldwarfsuperflare}, objects cooler than M7-M8 exhibit much less coronal X-ray and chromospheric H$\alpha$ emission \citep{Berger2010ultracool-dwarf-xray-halpha} than warmer spectral types, and this sharp difference suggests that auroral processes are the more significant magnetic activity for these cooler objects.

Most ultracool dwarfs observed to produce auroral emission also produce quiescent radio emission \citep{Pineda2017browndwarfaurorae}. Recent VLBI observations of one such ultracool dwarf have revealed this radio emission to be produced by trapped energetic particles in a large-scale dipolar magnetic field \citep{Kao2023radiationbelt}, analogous to the radiation belts of Jupiter \citep[e.g.][]{Zarka1998aurorareview}, further emphasizing the key role that large-scale magnetic fields play in defining the magnetic activity for these objects. 

However, large-scale fields are not ubiquitous at the end of the main sequence, and very different field strengths and topologies are observed even in coeval fully-convective dwarfs of equal mass \citep{Kochukhov2017}, with arguments put forward that the dynamo in rotating fully-convective objects may have two distinct stable states. 

At 2.7\,pc from Earth, the binary M dwarfs UV and BL Ceti are a well-characterized system for studying magnetospheres of fully-convective stars. With an orbital semimajor axis of 2.0584\,AU, they do not directly interact. UV Ceti nearly identically resembles its less-active companion BL Ceti in age, spectral type \citep[M6Ve and M5.5Ve,][]{Henry1994}, mass \citep[$0.1195 \pm 0.0043 M_{\odot}$ and $0.1225 \pm 0.0043 M_{\odot}$,][]{Kervella2016}, and rapid rotation period \citep[$0.2269 \pm 0.0005$\,day and $0.2432\pm0.0006$\,day,][]{Barnes2016}, but UV Ceti exhibits more frequent flares and much brighter persistent X ray and radio emission. 

Emission phenomena from UV Ceti include indicators of coronal and chromospheric activity typical of M dwarfs flare stars---frequent flares and saturated X-ray and H$\alpha$ emission---but also include auroral activity more similar to ultracool dwarfs and planetary magnetospheres. As UV Ceti is the archetype M dwarf flare star, the initial detection of periodic, highly circularly polarized radio emission, consistent with auroral ECME \citep{Lynch2017ECME, Zic2019, Villadsen2019}, therefore came as somewhat of a surprise. 

Different magnetic field configurations may interestingly underlie the difference in activity between UV Ceti and its twin BL Ceti. Zeeman Doppler Imaging confirmed that UV Ceti has a large-scale axisymmetric field and a rotationally-modulated field strength, while the magnetic field of BL Ceti is complex and non-axisymmetric \citep{Kochukhov2017}.  

 In addition to exhibiting frequent and intense flares, M dwarf flare stars such as UV Ceti exhibit persistent coronal emission which bears similarities (intense X-ray emission and a non-thermal radio spectrum)  to solar coronal emission during flares (see e.g.\ \citep[e.g.][and references therein]{Dulk1985}.  UV Ceti was one of the first main sequence stars detected at radio wavelengths, and the initial persistent detection of bright, non-thermal, coronal emission also came as a surprise since thermal Bremmsstrahlung emission dominates the solar corona \citep{Gary1981}. This persistent emission, and the more recent discovery that ultracool dwarf auroral emitters tend also to exhibit persistent non-thermal radio emission \citep[e.g.][]{Pineda2017browndwarfaurorae, Kao2023radiationbelt} highlight questions about the nature of the high-energy particle acceleration and particle-trapping required to power this persistent emission. 
 
Gyrosynchrotron, synchrotron, and coherent emission processes make radio wavelengths a window for observing magnetic activity and high-energy particle acceleration in stars. Solar flares accelerate particles to mildly relativistic energies, up to $\sim 1$\,MeV for electrons and 1\,GeV/nucleon for atomic nuclei, and these particles produce X-ray and non-thermal radio emission in the sun's magnetic field during the flare \citep[e.g.][]{Dulk1985}. Magnetized planets and compact stellar remnants (magnetic white dwarfs and pulsars) accelerate high-energy nonthermal particles in stable large-scale current configurations. 

 Radio observations over a wide range of wavelengths can delineate the interplay of coherent and incoherent, thermal and non-thermal, and coronal and auroral emission processes. We present an analysis of archival observations of UV Ceti with the Karl G.\ Jansky Very Large Array (VLA) at 1--40\,GHz and the Atacama Large Millimeter Array (ALMA) at 90--104\,GHz. The VLA observations we include in this analysis investigate auroral and coronal processes over a range of stellar rotational phases. Section 2 presents details of the VLA and ALMA observations, section 3 presents the results, section 4 discusses possible emission scenarios, and section 5 concludes.

\section{Methods} 

\subsection{Observations}
UV Ceti was observed by the VLA on January 8 2011 in three bands between 1 and 25\,GHz, and on January 28 2011 in five bands between 2 and 40 GHz (Project code 10C-210). For each VLA observation, UV Ceti was observed with each receiver sequentially, using all antennas available. The VLA was in B-North-C configuration for both observations.  UV Ceti was observed by ALMA on July 23 2014 in band 3, from 90 to 105 GHz (Project code 2012.1.00993.S). Table 1 summarizes the observations.  The total time elapsed between the start and finish of the set of observations, including time spent on calibrators, was 2 hours on January 8 2011 and 2.5 hours on January 28 2011. During the L band observation, a phase calibrator was observed for three minutes before beginning to observe UV Ceti. For the S band observation, a phase calibrator was observed for one minute at the beginning and again at the end of the observation. The C band observations were bracketed by 90 second and one minute phase calibrator observations on January 8 and January 28, respectively. At Ku band, the phase calibrator was observed for one minute every six minutes throughout the observation. At K and Ka bands, the phase calibrator was observed for one minute every three minutes. 
The Ku band and S band images had reduced sensitivity because these VLA observations were made at the end of a major upgrade to the array, and the S band and Ku band receivers were not yet available on many of the antennas. The ALMA observation consists of a single receiver band and thus its entire range of observed frequencies was observed simultaneously.  The ALMA observation included observations of a phase calibrator roughly once every seven minutes for 30 seconds at a time, and an atmosphere water vapor calibration once every 15 minutes.

\subsection{Calibration}
We used CASA  \citep[Common Astronomy Software Applications software package,][]{McMullin2007-CASA} to flag, calibrate, and image the VLA data, and to image the pipeline-calibrated data obtained from the ALMA archive. For the VLA observations, 3C48 served as both flux density calibrator and bandpass calibrator. J0132-1654 was used as the phase calibrator at all frequencies except for L band (2-4\,GHz), for which J0116-2052 was used.  For the ALMA observations, Uranus served as the flux density calibrator, J0132-1654 as the phase calibrator, and J0137-2430 as the bandpass calibrator. 

Typical absolute flux density calibration uncertainty for the VLA is 5\% at L--Ku bands and 10--15\% at K and Ka bands \citep{Perley2017}. 
For the L band phase calibrator J0116-2052, our flux density measurement is within 6\% of the VLA calibrator catalog value, consistent with the expected systematic uncertainty.  The phase calibrator for the other VLA bands, J0132-1654 appears to be several times brighter than the flux densities in the VLA calibrator catalog. Long term monitoring of J0132-1654 at Ku band with the OVRO 40m telescope \citep{Richards2011} shows that J0132-1654 underwent a large, slow brightening, and our flux density measurement at Ku band agrees with the OVRO measurement from the following day (Sebastian Kiehlmann, private communication).

\subsection{Imaging}
We imaged the visibility data using the CASA tclean routine. We used Briggs 0 weighting, except with L band for which we used Briggs -0.5 to better mitigate sidelobes from a bright source roughly 20 arcminutes to the south east of UV Ceti.  None of these archival observations include polarization calibrators, but since the VLA has left- and right-hand circularly polarized feeds, circular polarization (Stokes V) can be estimated without an additional calibrator.  We imaged UV Ceti in Stokes V for all the VLA observations, using imaging parameters that were otherwise similar to the Stokes I images. Only total intensity (Stokes I) images are possible for the ALMA data: ALMA's feeds are linearly polarized and only XX and YY correlation visibilities (no XY and YX correlations) were available in 2014 when the observation ocurred. 

We measured the flux density of UV Ceti and the unresolved calibrators two ways: first by fitting a Gaussian source to the cleaned image via the CASA Gaussian fit tool and obtaining a best fit flux density and statistical uncertainty from the fit, and secondly by measuring the peak pixel value on the source in an over-sampled image and the RMS in a region off the source.  These two methods provided similar results.  Flux density results are reported in figure \ref{fig:flux}. 

At L band, a bright source 19.7 arcminutes from UV Ceti near the edge of the primary beam
required self-calibration to reduce the contamination by its sidelobes, which otherwise would be inadequately removed by the CLEAN algorithm.  We imaged nearly the full primary beam at L band, using three terms in the Taylor expansion.  We used this first iteration image to create a model of the bright source, via the CLEAN algorithm, and then solved for a new phase-only calibration using that model.

\subsection{Light Curves}
We created light curves from the complex visibilities as follows. If the only source in the field of view is a point source at the phase center, then the real component of the visibility for any baseline is the flux density of the source, and the imaginary component is zero. We used the CASA task fixvis to move the phase center to the position of UV Ceti (since the source is in a binary system, the original phase center was near but not centered on UV Ceti). At L and S band, where other sources are in the field of view of the primary beam, we used the CLEAN algorithm to create a sky model of all of the bright sources in the field of view except UV Ceti, and then subtracted these model visibilities from the measurement set. Subtraction of background sources was not necessary at higher frequencies due to the narrower primary beam and the spectrum of the sources. After subtracting the sky model of the other sources from the visibilities, we averaged the visibilities over all channels, spectral windows, and baselines to create a time series of the flux density from UV Ceti. We obtained Stokes I and Stokes V time series from the separate time series made with visibilities from correlations of all of pairs of right hand circularly polarized feeds and separately all of the left hand circularly polarized feeds.  We searched for variability on a range of timescales in each observation. We use the imaginary components of the complex visibilities to estimate the uncertainties in the flux densities of the light curves.  For an image of a single point source at the phase center, the imaginary components of the complex visibilities will be pure noise with the same noise properties as the real visibilities. The standard deviation of the Stokes I time series computed from the imaginary components estimates the statistical uncertainty in the flux density. Not all bands and observations imaginary components consistent with zero mean, indicating a systematic uncertainty due either to contamination from other sources not full subtracted, or a small offset between the phase center and the position of the source. The statistical uncertainties dominate these systematic uncertainties in all observations except January 28 C band, for which they are comparable. Thus, the standard deviations of the Stokes I timeseries of imaginary components are used as representative errorbars in Figures \ref{fig:allIV}, \ref{fig:kflare}, and \ref{all}.  

We have not attempted to subtract BL Ceti from the visibilities, but we confirmed that the position of all flares is most consistent with the position of UV Ceti, as described in Section 2.5.

For additional confirmation of the slow variability observed with ALMA, we made single-scan images for each of the scans in the observation and confirmed from the point spread function in each single-scan image that the slow change was not due to a change in the quality of the phase calibration. Furthermore, we repeated the single-scan imaging for the phase calibrator and confirmed that the flux density from phase calibrator did not vary during this time. 

\subsection{Astrometry}

As mentioned above, UV Ceti and BL Ceti are a wide binary with a 26-year period. We determined the expected positions of UV Ceti and BL Ceti in January 2011 (the time of the VLA observations) and July 2014 (the time of the ALMA observation) as follows. We used the coordinates and proper motions of UV and BL Ceti in the GAIA DR3 catalog to determine the position of the system barycenter at each epoch.  We calculated the position angle and separation of the binary using masses and orbital parameters from \cite{Kervella2016} as input to the ephemeris calculator tool developed by Brian Workman\footnote{ Available from \url{https://www.saguaroastro.org/wp-content/sac-docs/ObservingDownloads/binaries_6th_Excel97.zip} as of February 3 2023}.   The expected angular separation between the two components of the binary was 2.1\,arcseconds in January 2011 and 2.3\,arcseconds in July 2014. The ALMA observation and the VLA observations at K band and Ka band have adequate resolution to resolve the binary, but the  K band observation on January 28 2011 is the only observation where both components of the binary are detected.

\begin{figure*}[ht!]
\includegraphics[width=\textwidth]{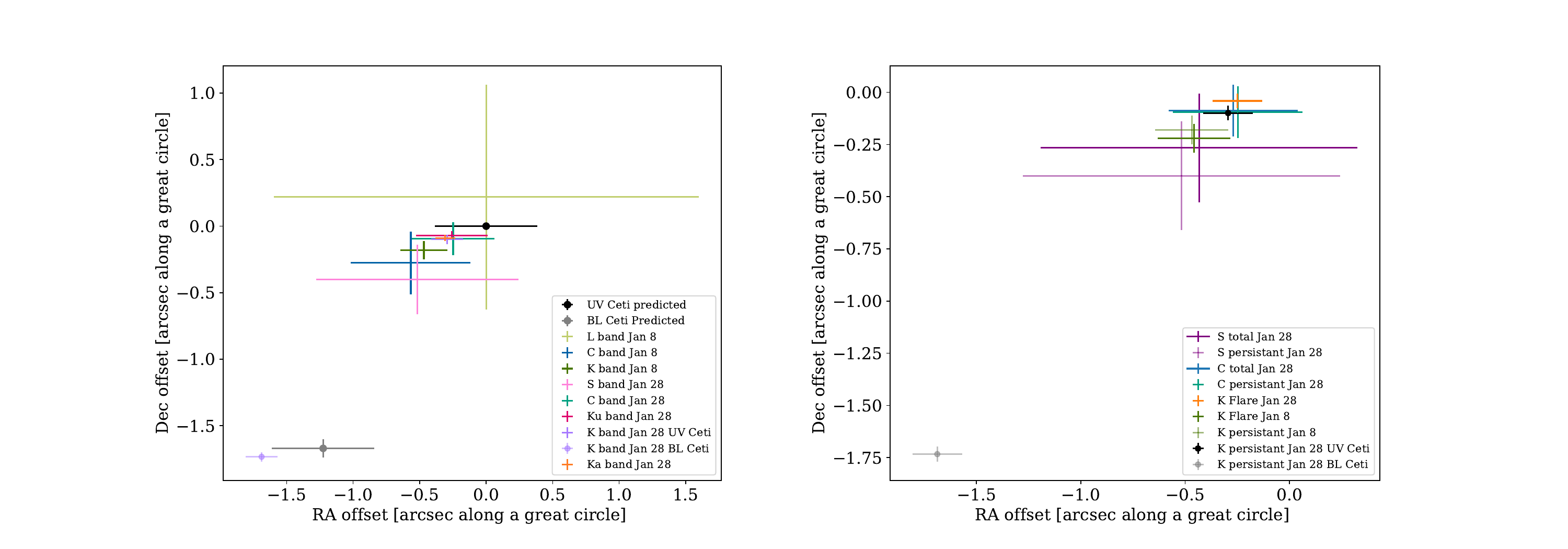}
\caption{Positions of the emission observed with the VLA, shown as an offset relative to the predicted position of UV Ceti. The offset is calculated as the observed position minus the expected predicted of UV Ceti. Both axes are in dimensions of arcseconds along a great circle; the RA offsets have been converted to seconds of degrees and multiplied by the cosine of the declination. The errorbars show the systematic uncertainties for each observation, since these are larger than the statistical uncertainties. \textbf{Left:} Persistent emission. \textbf{Right:} Variable emission compared to corresponding persistent emission. In this panel, offsets are still computed relative to the predicted position of UV Ceti, but for visual clarity only the observed positions of the two components are shown.}\label{fig:astrometry}
\end{figure*}

We measured the positions of the source(s) in all images by fitting a Gaussian source to the cleaned images via the CASA Gaussian fit tool. The best fit source positions relative to the predicted position of UV Ceti are shown in Figure \ref{fig:astrometry} for VLA observations.  The angular sizes (full width at half maximum, FWHM) of the synthesized beams are shown in Table 1.
At lower frequencies than K band, the synthesized beam is much wider than the binary separation. The statistical uncertainty on the centroid position of the Gaussian fit and the systematic astrometric uncertainties of the VLA, however, are much smaller than the synthesized beam, allowing the identification of the binary component responsible for the majority of the flux density. The systematic astrometric uncertainties of the VLA and ALMA are 10\% of the FWHM of the primary beam for both instruments, and this uncertainty is shown as the error bars in Figure \ref{fig:astrometry}, since it is larger than the statistical uncertainties. Note that the systematic offset will be different for each observation and band. The uncertainties in the predicted positions of the binary components due to the uncertainties in the GAIA positions and proper motions are less than a milliarcsecond, and thus negligible compared to the errors introduced by uncertainties in parameters in the orbital position angle calculation. We estimate the uncertainty in predicted positions by comparing the predictions to the observed positions in the image in which both components are detected. In Figure \ref{fig:astrometry}, the error bars on the predicted positions are the mean RA offset and mean declination offset from this comparision. Note that this method of estimating the uncertainty in the prediction introduces a roughly 0.1\,arcsecond systematic uncertainty to the entire plot (due to the systematic uncertainty in the observed K band position). For all of the VLA observations of persistent emission with a single source detected, that source lies significantly closer to UV Ceti than BL Ceti (Figure \ref{fig:astrometry}). Due to the larger beams at lower frequencies, a contribution from BL Ceti at L and S band cannot be ruled out. At C band on January 8 (the less time-variable C band observation) the centroid position suggests that the BL Ceti makes a non-negligible but lesser contribution to the total flux density, as is expected.

The self-calibration used at L band has the potential to introduce an astrometric offset.  We verified the L band astrometry by comparing the position of UV Ceti before and after the self-calibration, and by comparing the position of a nearby source in the final self-calibrated image to the VLASS catalog.  The shift in the measured position of UV Ceti before and after the self-calibration is 0.27\,arcseconds and the offset from the VLASS catalog is 1.3\,arcseconds, which are both smaller than the expected astrometric precision of 10\% of the synthesized beam.

In order to identify which component of the binary is responsible for the flares observed in the K band observations, for each observation we created a model of the persistent emission from imaging the timerange of the observation before and after the flare. We subtracted this model from the visibilities at the time of the flare, and then imaged the flare and measured its position as described above. For both flares, the flare position is consistent with the position of UV Ceti, within the uncertainties.

The C band and S band observations may not include any quiescent emission, and so for the purposes of identifying the source of the activity, the position of the source in an image of the full observation is compared to the position in an image of the least-active portion of the observation, rather than subtracting a model.  In both cases, the positions are consistent with UV Ceti being the source of the activity. In the S band observation where the persistent component suggests a contribution from BL Ceti, the source position moves closer to UV Ceti when the bright burst is included in the image.

The ALMA data is not included in Figure \ref{fig:astrometry} because the time between the VLA and ALMA observations is long enough for a non-negligible shift in the positions of UV and BL Ceti.  The best-fit position of the source observed with ALMA is $0.09\pm0.41$ arcseconds from the expected position of UV Ceti and $2.21\pm0.41$ arcseconds from the expected position of BL Ceti, thus confidently identifying the source as UV Ceti.  The uncertainty of $0.41$\,arcseconds is obtained from summing in quadrature the systematic uncertainty of the ALMA position and the uncertainty on the predicted position, estimated as described above.  Since the ALMA observation is nearer than the VLA observation to the GAIA epoch, this uncertainty may be an overestimate.

\begin{deluxetable*}{cccccccc}
\tablenum{1}
\tablecaption{Observations\label{tab:obs}}
\tablewidth{0pt}
\tablehead{
\colhead{Band} & \colhead{Array} & \colhead{Frequencies} & \colhead{Polarization} & \colhead{Duration on} &  \colhead{Duration on} &  \colhead{Duration on} & \colhead{Beam size} \\ 
\nocolhead{}& &  & & \colhead{Jan-8-2011 } & \colhead{Jan-28-2011}& \colhead{Jul-23-2014 } \\
\nocolhead{}& & \colhead{[GHz]} & & \colhead{[minutes]} & \colhead{[minutes]}& \colhead{[minutes]} & \colhead{[arcseconds]}\\
\vspace{-20pt}
}
\decimalcolnumbers
\startdata
L & VLA & 1--2 & Stokes I,V & 29 & -- & -- & 16.6$\times$8.8\\ 
S &VLA & 2--4  & Stokes I,V& -- & 20 & -- & 7.9$\times$2.7\\ 
C & VLA& 4-5 \& 7-8& Stokes I,V & 29 & 14 & --& 4.7$\times$2.5, 3.2$\times$1.3\\ 
Ku & VLA& 13--14 \& 16--17 & Stokes I,V& -- & 25 & -- & 2.8$\times$0.4\\ 
K &VLA & 19--20 \& 24--25 & Stokes I,V& 24 & 18 & -- & 1.8$\times$0.7, 1.2$\times$0.4\\ 
Ka & VLA& 30--31 \&  39--40 & Stokes I,V &  -- & 18 & -- & 0.8$\times$0.2\\ 
ALMA band 3 & ALMA & 90--94 \& 101--105& Stokes I & --& --& 37 & 1.3$\times$0.5\\ 
\enddata
\tablecomments{Summary of observations. Duration is the number of minutes spent observing UV Ceti. The beam size is the angular size of the full width at half maximum of the synthesized beam, in the dimensions of right ascension by declination. Where two beams are listed, these correspond to the January 8 and January 28 observations, respectively.}
\end{deluxetable*}

\section{Results}
\begin{figure*}[ht!]
\includegraphics[width=\textwidth]{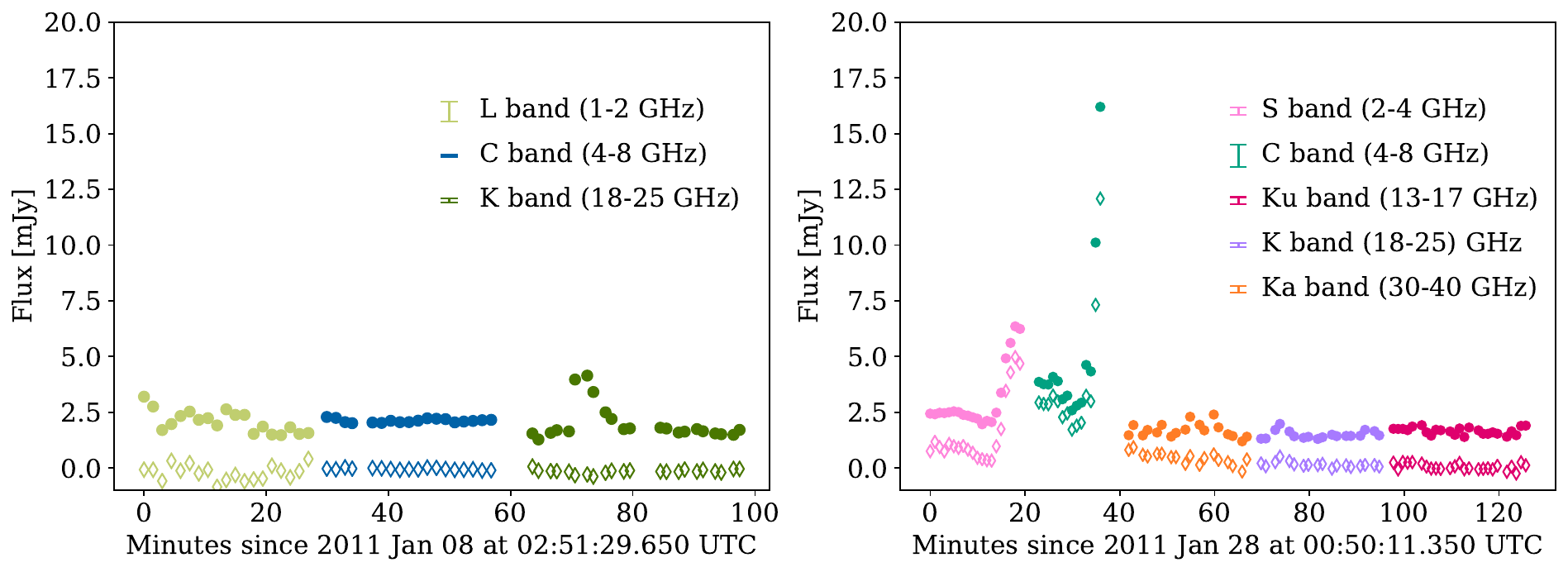}
\caption{Light curve of UV Ceti for all of the VLA observations (left: January 8 2011; right: January 28 2011). Stokes I flux density is shown with filled circles. Stokes V flux density is shown with open diamonds. Color indicates the receiver band. The measured flux density at each receiver band is shown, with no re-scaling to remove a spectral index. The time series are formed by moving the phase center to the position of UV Ceti and then averaging the visibilities over all baselines and frequency channels. The resulting light curves are binned to one data point per scan. Scans were one minute long for all except January 8 L Band and C Band, which were 90 seconds. Representative errorbars are shown in the figure legend, and are estimated as the standard deviation of the imaginary components of the complex visibilities, as described in section 2.4.  Gaps in the time series were times during which the phase calibrator was observed, with the exception of a few gaps where scans were flagged due to contamination by RFI. For the L band and S band data, other bright sources in the field of view of the primary beam have been subtracted from the visibilities prior to calculating the light curve. This was not necessary at higher frequencies due to the narrower primary beam and the spectrum of the background sources. \label{fig:allIV}}
\end{figure*}

We detect UV Ceti at all epochs and observing bands. It exhibits persistent, slowly varying, circularly polarized emission as well as flares (Figure \ref{fig:allIV}). BL Ceti is only detected at K band on January 28 2011, and marginally detected at K band on January 8 2011. UV Ceti and BL Ceti were separated by 2.1 arcseconds in January 2011 and 2.3 arcseconds in July 2014 and thus, had BL Ceti been brighter, it would have been detected as a second source at least marginally resolved from UV Ceti in all the observations at frequencies above 10 GHz. Below 10 GHz, the separation of the components of the binary is smaller than the half-power width of the synthesized beam, but since the uncertainty on the position of the centroid of a point source is proportional to the width of the point spread function divided by the signal-to-noise ratio, we can fit the source position with sufficient precision to determine that UV Ceti, and not BL Ceti, dominates the observed flux density at all observed frequencies.

\subsection{Spectrum}
\begin{figure}[ht!]
\includegraphics[width=1.0\columnwidth]{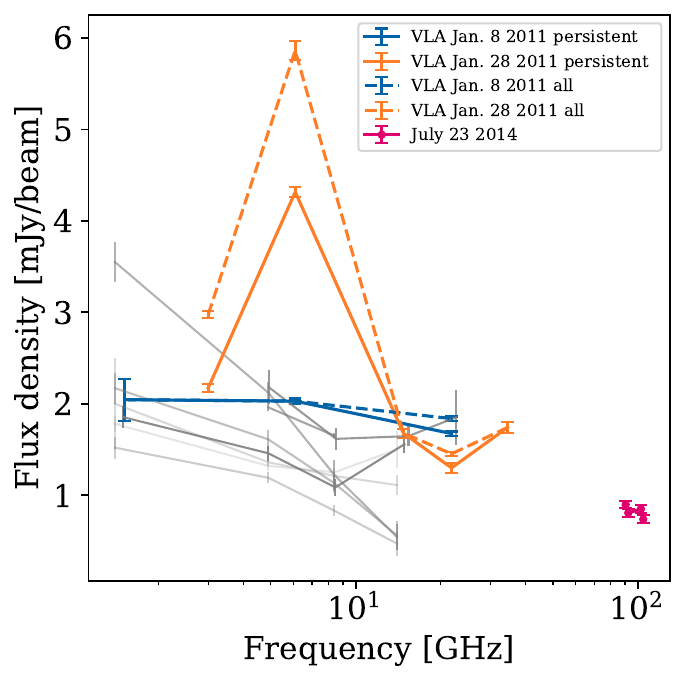}
\caption{Flux density vs frequency measured with the VLA and ALMA, with error bars showing statistical uncertainties.  Typical systematic uncertainties for the VLA are on the order of 10--15\%. Color indicates the different observing epochs. Lines connect the measurements at each band to make the groups of observations for each day easier to see. For the VLA observations, dashed lines, labelled `all', show the flux density at each band averaged over the full length of the observation. The solid lines, labelled `persistent', show the flux density averaged over the observation excluding times of flare or burst events. For the S and C band data, the large bursts have been excluded from the points labelled 'persistent', although the entire observation may be part of one extended burst.  For the VLA data the flux density shown is obtained from the peak pixel in the image of the source, while for the ALMA data it is the peak of a Gaussian fit to the source. The ALMA data is plotted with one point per spectral window. For comparison to other observations of UV Ceti in the literature, the gray lines show eight VLA observations from a five-year period 1987--1992, compiled in \cite{gudelbenz1996radiospectra} and  \citep{Gudel1989}. For legibility, these observations are not itemized in the legend. From darkest to lightest, the gray-scale corresponds to October 1987 (two observations), December 1990, June 1991, July 1991 (two observations), and January 1992 (two observations).
\label{fig:flux}}
\end{figure}

Figure \ref{fig:allIV} shows the left-hand and right-hand circularly polarized light curves of all the VLA observations.  Figure \ref{fig:flux} shows the average flux density of UV Ceti measured in each of the VLA and ALMA observations, plotted against the frequency of the band center of each observation.  On January 8 2011, the flux density from UV Ceti was fairly similar at each frequency band observed. On January 28 2011, UV Ceti was brightest during the C and S band observations. Since UV Ceti rotates with a period of 5.45 hours \citep{Barnes2016}, as well as flaring frequently, it is important to emphasize that the VLA observations in each band were conducted sequentially, not simultaneously, during each observation day. Figure \ref{fig:flux} does not report instantaneous spectra, rather the flux density measurements in each band are displayed on one plot for the purpose of providing a summary. In fact, the baseline component of the flux density in Figure \ref{fig:allIV} suggests that a persistent component of the emission from UV Ceti may have a fairly flat spectrum. For comparison, Figure \ref{fig:flux} also shows eight other VLA observations, from \cite{gudelbenz1996radiospectra} and \cite{Gudel1989}.  None are simultaneous measurements across the bands observed, and thus have the same limitation in interpreting the measurements as a single spectrum, although all were reported with an attempt to exclude flares. They are presented here to give a sense of the varied emission UV Ceti displays. The convex shape of the January 8 2011 plot has analogs in the earlier observations, as does the the up-turn at the higher VLA frequencies on January 28 2011 (although the earlier observations do not extend to such high frequencies). The C band emission on January 28 2011 is much brighter than the other examples, which further suggests that the even the apparently persistent component of the emission throughout this observation (with the bright flare removed) is not truly quiescent.

At Ka band, the upper and lower side bands are separated by nearly 10\,GHz (the lower sideband is 29.487--30.511\,GHz and the upper sideband is 38.487--39.511\,GHz). Since the two sidebands are observed simultaneously, we can estimate a spectral index by imaging the two sidebands separately. The Ka band spectrum rises to higher frequencies, with a spectral index of $0.63 \pm 0.27$.

Millimeter emission is confidently detected on July 23 2014 with ALMA but is significantly fainter than all of the VLA observations (Figure \ref{fig:flux}).   The millimeter emission has a spectral index $-0.81\pm0.36$ (with the convention that the spectral index $\alpha$ is defined such that at frequency $\nu$ the flux density $F$ is proportional to $\nu^\alpha$), measured by separately by imaging each of the four spectral windows in the ALMA observation and then fitting a power law to the measured flux density vs frequency.

\begin{figure}[ht!]
\includegraphics[width=\columnwidth]{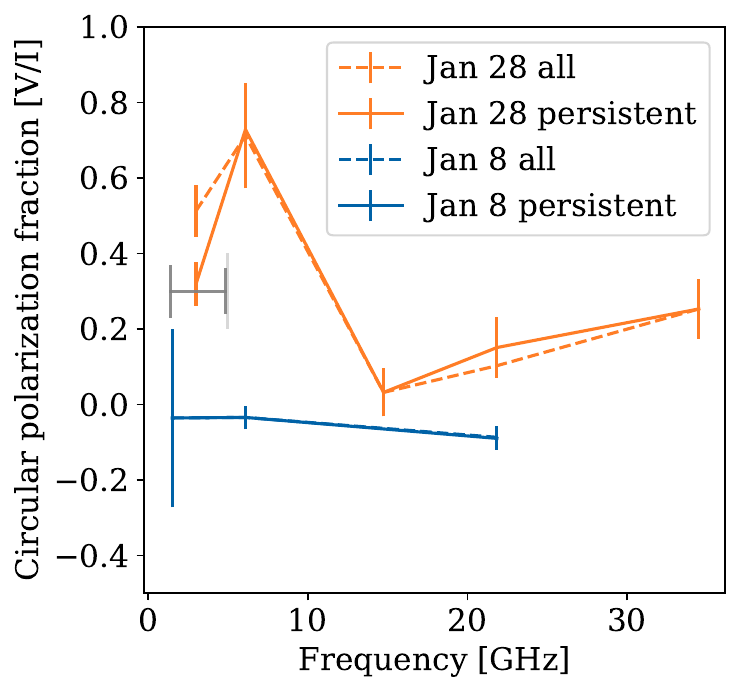}
\caption{Circular polarization fraction measured in the VLA observations, measured as flux density in Stokes V images divided by flux density in Stokes I images, with errorbars showing statistical uncertainties computed by summing the uncertainties of the Stokes I and Stokes V images in quadrature.  In images without a significant detection, we use the value in the Stokes V image at the position of UV Ceti determined from the corresponding Stokes I image.  Color indicates the different observing epochs. Lines connect the measurements at each band to make the groups of observations for each day easier to see. For the VLA observations, dashed lines, labelled `all', show the average over the full length of the observation. The solid lines, labelled `persistent', show the average over the observation excluding times of flare or burst events. For comparison, observations from \cite{Gudel1989} and \cite{Linsky1983} are shown in dark and light gray, respectively. }\label{fig:stokesv}
\end{figure}

Figure \ref{fig:stokesv} shows the circularly polarized fraction of flux density from UV Ceti (Stokes V / Stokes I). On January 8 2011, UV Ceti is unpolarized or weakly left-hand circularly polarized depending on the band. On January 28 2011, UV Ceti is nearly completely circularly polarized at 6\,GHz and has a significant component of right-hand circularly polarized emission at all frequencies observed except Ku band, which is nearly unpolarized. As with Figure \ref{fig:flux}, this plot of circular polarization fraction in each band is meant as a summary and does not necessarily imply spectral behavior, due to the non-simultaneous observations. 

Past observations of UV Ceti have ranged from unpolarized to 100\% circularly polarized, and when significant polarization is detected from UV Ceti it is typically but not always right-hand circularly polarized   \citep[e.g.][]{Gudel,Villadsen2019}.  In Figure \ref{fig:stokesv}, positive values indicate right-hand circular polarization and negative values indicate left-hand circular polarization.
Most of the observations from \cite{gudelbenz1996radiospectra}and \cite{Gudel1989} included for comparison in the stokes I plot in Figure \ref{fig:flux} do not quantitatively report polarization measurements: some qualitatively report a non-detection and for some cases polarization is not mentioned.  For the observation where a significant polarization is detected \citep{Gudel1989}, we include that observation as the dark gray points in Figure \ref{fig:stokesv} and note that it is similar to the level of polarization in the January 28 persistent emission. Since \citep{Gudel1989} do not report an uncertainty on their observed polarization fraction, we plot error bars based on propagating their reported Stokes I uncertainties. We additionally show, in light gray, and example from \citep{Linsky1983}. They present four observations of circularly polarized emission from UV Ceti at 5\,GHz from 1980-1981. Since all four observations are consistent with each other within their uncertainties, we plot only the most sensitive of their observations.

\subsection{Temporal Variability: VLA}
In the January 28 observation we detect intense, highly circularly polarized brightenings at C band and S band (see Figure \ref{fig:allIV}).  Both light curves are strongly circularly polarized both before and during the brightening events. 

\begin{figure*}[ht!]
\includegraphics[width=\textwidth]{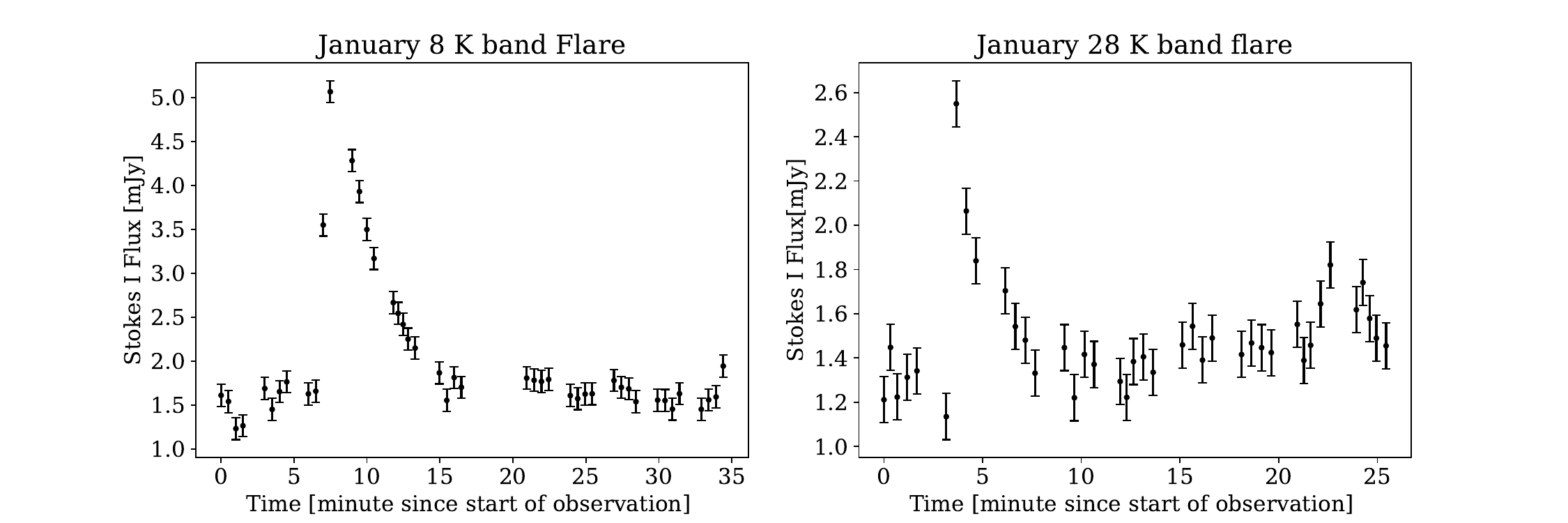}
\caption{Left: Stokes I flux density density vs time for the K band January 28 observation. Right: Stokes I flux density density vs time for the K band January 28 observation. In both panels, the time series are smoothed to 20\,s time resolution. Gaps are due to calibration scans and flagged data. The errorbars represent the statistical uncertainty, estimated as described in section 2.4.  \label{fig:kflare}}
\end{figure*}

At K band, we detect two flares--- one from each day. The flares rise sharply and decay approximately exponentially.  The Stokes I time series of each flare are shown in figure \ref{fig:kflare}.  Both flares have a spectrum rising to high frequencies, compared to flatter non-flare spectra.  During the January 8 flare, the K band emission from UV Ceti has a spectral index $0.56\pm0.17$ compared to $0.008 \pm 0.17$ during the nonflare time. If the average flux density from the non-flare part of the observation is subtracted from the total emission during the flare, then the spectral index becomes $0.99 \pm0.32$. 
The emission remains 7--9\% left-hand circularly polarized throughout the observation: during the non-flare time, the ratio of the Stokes V to Stokes I flux density, $V/I$, is $-0.09\pm0.01$. During the flare, $V/I$ is  $-0.07\pm0.02$ without subtracting the persistent component, and is $-0.052 \pm 0.037$ with the persistent component subtracted. These polarization fractions are consistent with no change during the flare. 

On January 28, the spectral index of $0.31 \pm 0.30$ for the persistent emission steepens to $0.69 \pm0.32$ during the flare, and the flare flux density after subtracting the persistent component has a spectral index of $1.4\pm1.28$. The fraction of circular polarization increases from $0.15 \pm 0.05$ in the non-flare time to $0.25 \pm 0.05$ during the flare, and the flare emission with the non-flare average subtracted is even more strongly circularly polarized with $V/I = 0.44 \pm 0.15 $. Additionally, on January 28, the baseline non-flare flux density shows a slow brightening during the observation.

The above estimates of the spectral index were obtained by imaging the upper and lower side bands separately for the time range of the flare. The uncertainty on each flux density measurement is obtained from the off-source RMS in each image, and these uncertainties are propagated by Monte Carlo process to constrain the uncertainty on the spectral index. The quoted uncertainties $V/I$ are estimated by taking the uncertainties from the off-source rms in each of the relevant images and analytically propagating the uncertainty through the polarization ratio calculation. 

\subsection{Temporal Variability: ALMA}
\begin{figure}[ht!]
\includegraphics[width=\columnwidth]{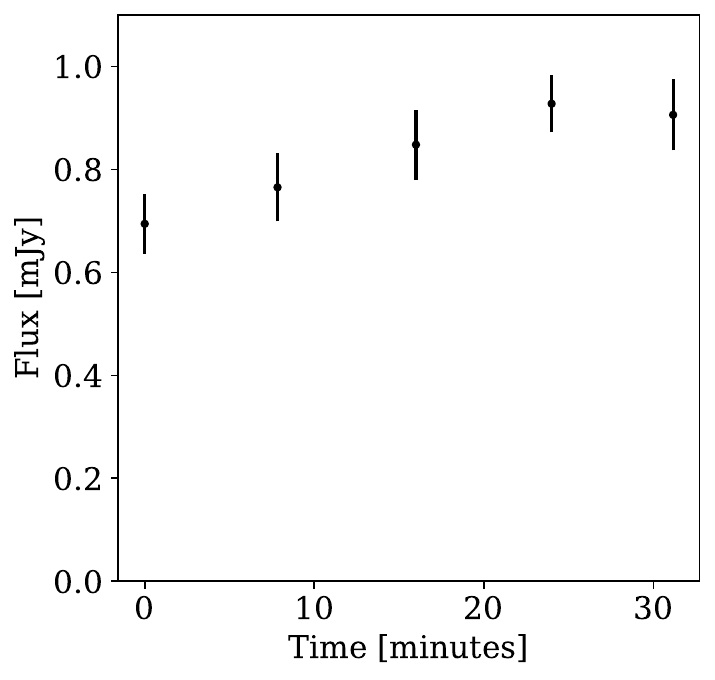}
\caption{Flux density vs time during the ALMA observation. The time is displayed in minutes since the start of the observation. The flux density values are the peak pixel value of an oversampled image, with one image made per correlator scan. Each of the first four scans is 6 minutes and 51.5 seconds long. The gaps between the scans are 30--45 second calibration observations. The final scan is five minutes and 16 seconds long. The error bars are the RMS in an off-source region of the image and are consistent with the expectation for thermal noise.\label{fig:almalc}}
\end{figure}

UV Ceti gradually brightened by 36\% during the ALMA observation, as shown by Figure \ref{fig:almalc}. The light curve in Figure \ref{fig:almalc} was made by imaging each of the five correlator scans separately, since there were fewer scans than in the VLA observations. The light curve made by averaging the visibilities with UV Ceti at the phase center shows the same brightening. The error bars in Figure \ref{fig:almalc} indicate statistical uncertainty due to thermal noise. These uncertainties are measured by computing the RMS of an off-source region of each image and were confirmed to be consistent with the expected sensitivity of this ALMA observation.  Within the uncertainties, the measurement from the final scan is consistent with a continuation of the same gradual brightening and also consistent with the flux density levelling off after the 36\% brightening.

We confirmed that the observed slow brightening is not caused by a change in the phase calibration in two ways. 
First, we confirmed that the flux density of the phase calibrator varies by only a few percent during this time. Second, if the phase calibrator is too far from UV Ceti, the quality of the phase calibration could change over time. We excluded this possibility by making a series of circular regions (in CASA region format) around UV Ceti, with increasing radius, and using the CASA imstat function to measure the integrated and the peak flux density in each region.  If the apparent brightening were due to phase errors causing an underestimate of the flux density in earlier scans, then those scans would have more flux spread over a larger solid angle, which was not the case. Thus, UV Ceti's apparent brightening is likely intrinsic. 

Using a higher time-resolution light curve computed by averaging the visibilities with UV Cet at the phase center, we searched for fast flares similar to the one-second millimeter flare from the M dwarf Proxima Centauri \citep{macgregor2021} and found none.

\section{Discussion}

In this section we discuss the possible emission mechanisms responsible for the combination of fast and slow phenomena we observed from UV Ceti. UV Ceti varies on such a range of timescales that ``quiescent'' may be a misnomer for any of the emission. Nontheless, time variability when clearly present, as well as brightness temperature, polarization and spectral index all inform our discussion of possible emission mechanisms. This section is organized in the following subsections: flares, auroral emission, and persistent emission. 

\subsection{Flares}

The short flares observed at K band have the fast-rising exponential-decaying light curves of flares consistent with magnetic reconnection events.  The 5--45\% degree of circular polarization, and steepening flare spectra (spectral indices of $0.99\pm0.32$ during the January 8 and $0.69\pm0.3$ during the January 28 flare) are also consistent with gyrosynchrotron emission.  Figure \ref{fig:harmonics} shows the field strengths required for the range of harmonics of the cyclotron frequency that occur for gyrosynchrotron emission. The observed frequencies are consistent with emission regions of hundreds of Gauss, well above the surface in the corona of UV Ceti.

 Optically thick gyrosynchrotron emission from isotropic electrons is mildly to moderately circularly polarized in the sense of the plasma O-mode, whereas optically thin gyrosynchrotron emission is polarized in the sense of the X-mode---the emission coefficient is higher for the X-mode but the source function is higher for the O-mode, see e.g. \cite{Dulk1979}.  The spectral indices of both flares rise to high frequencies, ruling out optically thin emission from a population of electrons with isotropic velocities. As discussed in \cite{Villadsen2019}, the north magnetic pole of UV Ceti remains in view throughout its rotation period, and thus O-mode emission averaged over the viewable hemisphere at any time in UV Ceti's rotation corresponds to left-hand circular polarization at Earth, and X-mode corresponds to right-hand circular polarization.  The K band flare emission on January 8 was left-hand circular polarized, as was the persistent K band emission during the observation. Thus, the January 8 K band emission is consistent with O-mode emission, if the magnetic field at the location of the emission was similar to the average field in the viewable hemisphere. On January 28, however, the K band emission (both the flare emission and the persistant component) was right-hand circular polarized, with the flare emission significantly more strongly polarized (40\%) than the persistent emission. This right-hand circular polarized emission would be consistent with X-mode, and thereby not consistent with optically thick gyrosynchrotron emission, if the field at the location of the emission source is in the same direction as the hemisphere average. Since flares associated with magnetic reconnection events typically arise from small-scale magnetic field features where the local field may not point in the same direction of the hemisphere average, it is difficult to use the sense of the circular polarization alone to definitively rule out or confirm the emission mechanism of short duration flares. In summary, the January 8 flare is consistent with gyrosynchrotron emission in a region aligned with a magnetic field aligned with the large-scale field, whereas a gyrosynchrotron explanation for the January 28 flare requires a more complex field structure or a more complex electron distribution.

The unknown size of the emitting region makes the brightness temperature of the flares difficult to estimate since the time resolution of the VLA observations is longer than the light crossing time across the stellar disk.  If the emitting region were the entire size of the stellar photosphere disk, the peak brightness temperature would be $8\times10^7$\,K and $4\times10^7$\,K for January 8 and 28, respectively. These brightness temperatures are an order of magnitude higher than coronal temperatures inferred from X-ray emission \citep[see table 2 in][]{Villadsen2019}, and may be much higher for smaller source regions.  Thus, thermal gyroresonance emission can be ruled out. 

\subsection{Auroral emission}
\cite{Villadsen2019} and \cite{Zic2019} observed periodic electron cyclotron maser emission from UV Ceti, which is the same process involved in auroral radio emission observed in Jupiter and ultracool dwarfs \citep[e.g.][]{Zarka1998aurorareview, hallinan2015aurora}. \cite{Villadsen2019} conclude that the coherent right-hand circularly polarized bursts they observed from UV Ceti are electron cyclotron maser emission because the bursts are polarized in the sense of the X mode, while plasma emission would be polarized in the sense of the O mode. The bursts we observe at C band and S band are polarized with the same sense as the bursts observed by \cite{Villadsen2019} and \cite{bastian2022} and thus consistent with being part of the same periodic electron cyclotron maser emission.  The maximum brightness temperature reaches $5.5\times 10^9$\,K for a source size of the stellar disk, which is likely a lower limit of the actual brightness temperature.

\begin{figure}[ht!]
\includegraphics[width=\columnwidth]{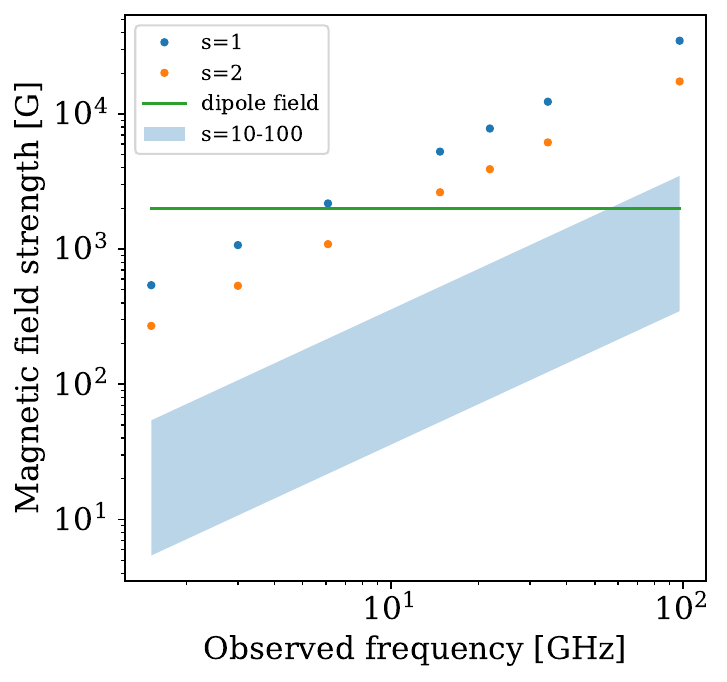}
\caption{Magnetic field strengths required for emission at the observed frequency to be the specified harmonic $s$ of the cyclotron frequency. Fields for the fundamental and second harmonics ($s=1$ and $s=2$) are plotted for the frequencies of the band centers of each of VLA and ALMA observations. The blue shaded region encompasses the range of harmonics applicable to gyrosynchrotron emission. The green line marks the polar surface field strength of the dipole component of UV Ceti's magnetic field. \label{fig:harmonics}}
\end{figure}

The observed frequencies of the ECM emission imply that the emission region spans field strengths of 0.5--1\,kG if the emission is at the second harmonic of the cyclotron frequency, and 1-2\,kG if it is at the first harmonic (see Figure \ref{fig:harmonics}).  The magnetic field of UV Ceti has a strong dipole component (field strength 2\,kG at the poles) as well as small-scale structure \citep{Kochukhov2017}. Considering only the dipole component, the field at radial distance $r$ can be approximated as $B(r)=2\,kG(r/r_*)^{-3}$ above the pole and a factor of two lower above the equator, where $r$ is measured from the center of the star and $r_*$ is the stellar radius measured by optical interferometry to be $0.159\pm 0.006$ solar radii \citep{Kervella2016}. 

In the dipole field, field strengths of 0.5--1\,kG occur at heights of 0.2--0.6 stellar radii above the surface of the star in the polar region and occur over the equator from 0.3 stellar radii down to the surface. If the emission is at the second harmonic, the required field strengths of 1-2\,kG occur from 0.2 stellar radii down to the polar surface, and do not occur above the equatorial surface unless higher order components in the magnetic field are considered.

\subsection{Persistent or slowly-varying emission}
This section discusses the persistent, slowly-varying emission phenomena observed, including a potential observation of a variation in circular polarization with rotational phase, and a gradual brightening in the ALMA observation.  

\begin{figure*}[ht!]
\includegraphics[width=\textwidth]{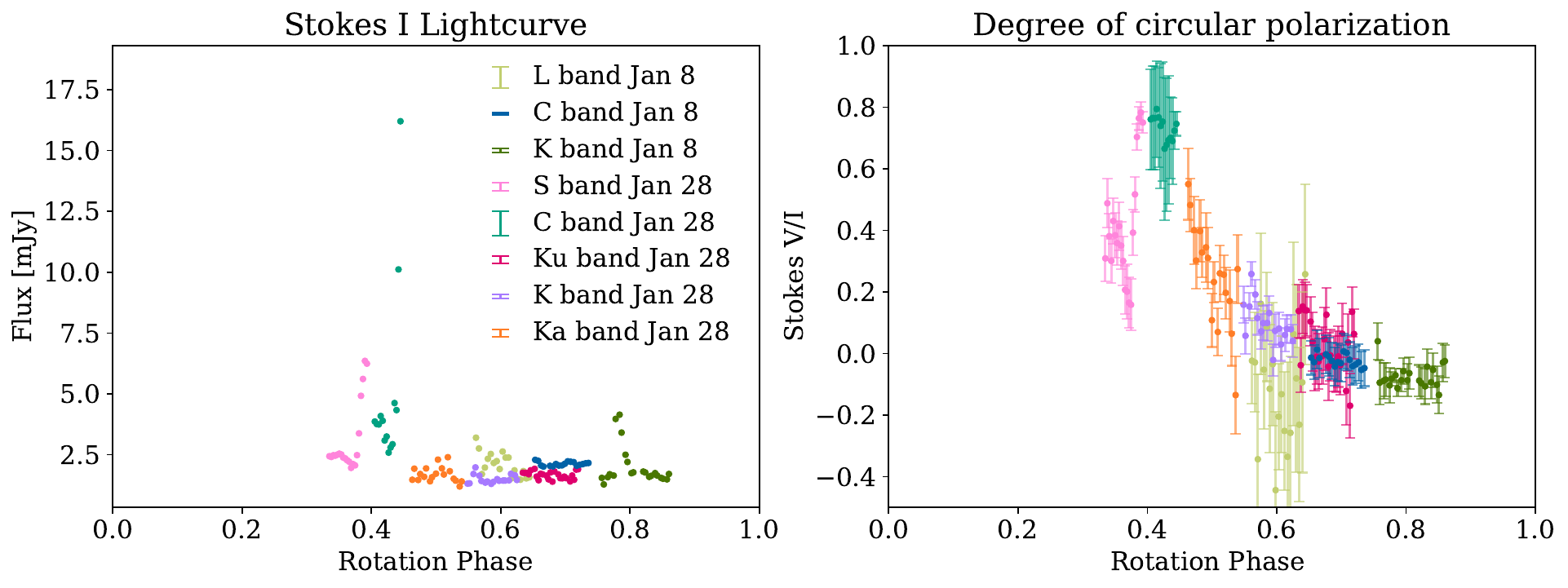}

\caption{VLA light curves for all observations, plotted as a function of rotational phase. The zero phase is arbitrary but consistent for all observations. light curves are binned to one data point per scan. Scans were one minute long for all except January 8 L Band and C Band, which were 90 seconds.   Left: Stokes I light curves. Representative errorbars are shown in the figure legend, and are estimated from the imaginary components of the complex visibilities as described in section 2.4. Right: Degree of circular polarization. The colors in the legend from the right panel apply to the left panel, but the errorbars are displayed for each point because the uncertainty on the polarization fraction, $\sigma_\mathrm{v/I}$, is obtained by propagating the uncertainty on the flux density, such that $\sigma_\mathrm{v/I} = \frac{\sigma}{f_I}\sqrt{1+\frac{f_v}{f_I}}$, where $f_v$ and $f_I$ are the Stokes V and Stokes I flux densities, respectively.  \label{all}}
\end{figure*}
\subsubsection{Slowly-varying circular polarization fraction}
No flares were found at Ka band, but there is a gradual decline in the degree of polarization from 40\% circularly polarized to 10\% circularly polarized. This observation occurred following the intensely circularly polarized C band observation.
 January 8 2011 and January 28 2011 are close enough in time to estimate the relative stellar rotational phase between all of the VLA observations. Figure \ref{all} plots light curves from all of the VLA observations at all frequencies vs relative rotational phase. As with figure \ref{fig:allIV}, the observed flux density in each band is plotted without any scaling by the spectral index. The polarization fraction of the combined lightcurve intensely increases during the C band and S band January 28 observations, with all the other observations lying on the approximately exponentially-declining tail.

The Ka band observation shows a slow decline in the degree of polarization from 40\% to 10\% circular polarization over the duration of the observation, which occurred immediately after the lower-frequency ECME observation. Comparing the circular polarization fraction to stellar rotation phase (Figure \ref{all}) shows a peak at the ECME followed by a slow decay. All the observations, regardless of frequency and observing day fall on this curve.  Two interpretations are possible.  The higher-frequency observations on January 28 may have coincided with an unrelated flare. Interestingly, the observations from January 8 lie on approximately the same curve, although the lightcurve has nearly flattened by that portion of rotation phase. It is unfortunate that the only observations overlapping between days both suffer from low signal to noise (due to fewer available antennas and data lost to RFI). The other possibility is that the emission is rotationally modulated at all frequencies. Simultaneous observations of UV Ceti at multiple frequencies over its full rotation would clarify this picture.

If the Ka band emission were electron cyclotron maser emission, the implied magnetic field strengths exceed 10\,kG for the fundamental frequency and 5\,kG for the first harmonic, for an observing frequency of 30\,GHz. These field strengths exceed the surface strength of the 2\,kG dipole component. 

In Jupiter's magnetosphere, electron cyclotron maser emission due to plasma from Io results from stable large-scale currents in a mostly dipolar field, but higher order terms in the magnetic field are required in order to produce field strengths that account for the highest frequencies at which the Jovian Io decametric radiation is observed \citep{queinnec1998jupiterdam,connerney1992-jupiter-magnetic-field}.
UV Ceti does have significant small-scale magnetic fields in addition to its strong dipole component, with a mean surface field strength of 6.7\,kG, suggesting that emission at low harmonics of the cyclotron frequency may be possible even at the highest frequencies in the VLA observation. 

Furthermore, at 30\,GHz, the requirement that the emission frequency sufficiently exceed the plasma frequency places less-stringent requirements on the density of the emission region compared to ECM emission at S band and C band. 30\,GHz emission could easily propagate through typical M dwarf coronal densities of $10^{10.5} \rm{cm}^{-3}$ \citep[see list of M dwarf coronal densities compiled in ][]{Villadsen2019}.  

If the emission at Ka band is not related to the periodic ECME, gyroresonance and gyrosynchrotron emission could be considered. Gyroresonance emission occurs at the cyclotron frequency and its first few harmonics. As with the ECME scenario, gyroresonance emission at Ka band would thus require field strengths corresponding to the small-scale field structures near the surface. Gyroresonance emission can be circularly polarized, but since gyroresonance emission is not beamed, the observed change in polarization would require the majority of the polarized emission to be confined in a region less than 10\% of the stellar surface, in order to rotate out of view over the course of the 18-minute observation. 

Flares cause time-variable gyrosynchrotron emission, although these flares often accelerate electrons with a power law distribution, which produce typically only moderately circularly polarized emission.
Gyrosynchrotron emission from a thermal population of electrons can be highly circularly polarized ($>$60\%), with a strong, non-monotonic frequency dependence in the polarization fraction (Golay et al.\ 2023 submitted), although explaining the change in polarization over 18 minutes is more difficult for a thermal population. Furthermore, the 10 million Kelvin brightness temperature is somewhat high for a coronal temperature (unless the emission region is several times the size of the stellar disk), and the right-hand circular polarization of the emission corresponds to X-mode for the average field of the viewable hemisphere, whereas optically thick gyroresonance or gyrosynchrotron emission is polarized in the sense of the O-mode.

Flares from the active corona may accelerate electrons that become trapped in radiation belts, making anisotropic pitch angle distributions important to consider.
The presence of a strong dipole field \citep{Kochukhov2017} suggests that radiation belts similar to those recently observed in an ultracool dwarf \citep{Kao2023radiationbelt} may be an important reservoir of radiating electrons.  Fully modelling the spectrum of UV Ceti may require a 3D radiative transfer simulation such as that used by \cite{Morris1990}, who modelled emission from a population of electrons with a thermal component and a non-thermal tail, trapped in radiation belts around a star, and used their model to describe the active member of an observed RS CVN type binary system. They observed a change in polarization sense with frequency, but they did not consider pitch angle anisotropies.  If the emission observed from UV Ceti at high frequencies is gyrosynchrotron, the emitting electrons may have an anisotropic pitch angle distribution, as has been used to explain radio emission from Jupiter's radiation belts \cite{depater1980-jupiter-thesis}. 

\subsubsection{Millimeter emission}
At millimeter frequencies, the flux density implies a brightness temperature of $6.5\times 10^5$\,K for an emitting region the angular size of the photosphere. The quiescent sun becomes optically thick to millimeter emission at the transition region at the top of the photosphere \citep[e.g.][]{Alissandrakis2022mm_sun_review}, which is much cooler for UV Ceti, and thus the observed persistent millimeter emission is too bright to have an origin analogous to millimeter emission from the quiet sun.  Gyroresonance emission is not likely at millimeter wavelengths because the implied field strengths are too large. If the emission is gyrosynchrotron emission, the field strengths implied are a few hundred to a few thousand Gauss depending on the harmonic number, which is consistent with the field expected for the stellar surface or low corona, and large emitting region-filling factors are required. 

If the emission emanates from an extended large-scale magnetosphere, the brightness temperature could be 100 times lower (for a magnetosphere $\sim10$ stellar radii in diameter). Emission regions of size scales 1--10 stellar radii correspond to brightness temperatures of $6.5\times 10^3$ -- $6.5\times 10^5$\,K which are within the 3--6\,MK range of coronal temperatures estimated by \cite{Audard2003UVcetiXray}.  However, depending on where exactly the coronal temperature lies in that range and how extended the millimeter emission is, the millimeter emission may be required to be optically thin.

The millimeter emission is slowly variable, brightening by 36\% over 30 minutes.  If this variability is caused by a brighter region of the star rotating into view on or near the stellar surface, then the minimum brightness temperature of that region, $T_{\rm{region}}$, compared to the brightness temperature $T_{\rm{baseline}}$ prior to the brightening is $$ \frac{T_{\rm{region}}}{T_{\rm{baseline}}} = \frac{\Delta f}{f}\frac{P}{t}$$  

where $P$ is the rotational period (0.2269 days), $t$ is the timescale of the brightening (30 minutes), and $\Delta f/ f$ is the fractional change in flux density (36\%).  Thus, using rotational modulation to explain the 36\% brightening over 30 minutes would require a region with a brightness temperature at least four times the brightness temperature of the rest of the star to rotate into view.  Alternatively, the variability could be intrinsic and related to a slow flare. 

No fast flares similar to that observed for Proxima Centauri \citep{macgregor2021} were detected. However, if UV Ceti exhibited similar millimeter flares, these fast flares would occur too infrequently to expect a detection in an observation of the length we have in the ALMA archive.  Millimeter flares on M dwarfs have been detected with timescales ranging from seconds \citep{macgregor2018-mm-flare,macgregor2020-mdwarf-mm,macgregor2021,ward2022-mm-prox-cen} to tens of minutes or longer\footnote{Some observations ended before the end of the flare.} \citep{Guns2021, naess2021-act-mm-transients}.  It is possible that the millimeter variability we observe on UV Ceti is part of a gradual flare.  Future longer observations could identify flares or rotational variability, and polarimetry would help identify the emission mechanism.

\subsubsection{Brightness Temperature as a function of frequency}
\begin{figure}[ht!]
\includegraphics[width=\columnwidth]{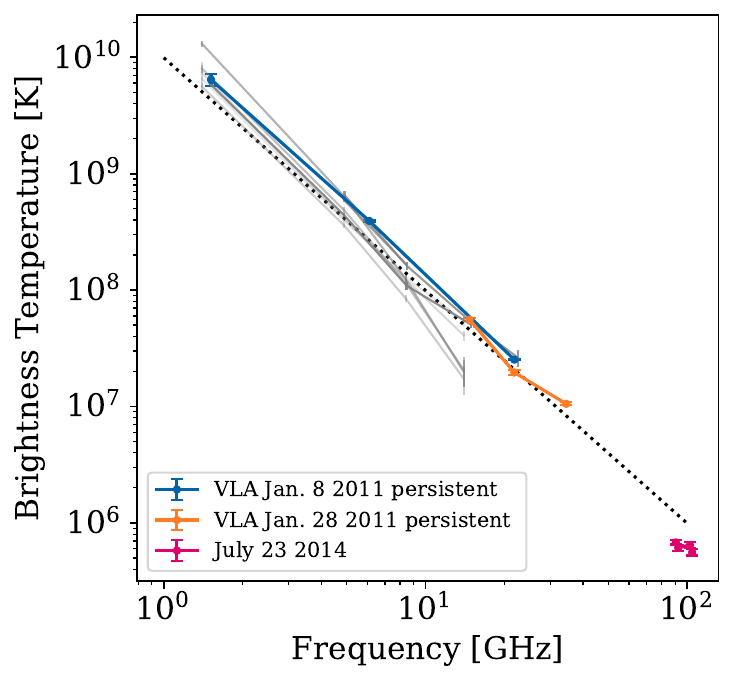}
\caption{Brightness temperature for the persistent component of emission at all bands for which persistent emission was observed (the January 28 S band and C band data is excluded as the ECME burst dominated the observation). The brightness temperature is calculated for the angular size of the photospheric disk, and would be lower if the emitting region encompasses an extended magnetosphere. Color indicates the different observing epochs. Lines connect the measurements at each band to make the groups of observations for each day easier to see. For the ALMA data, one point is plotted for each spectral window. Errorbars are calculated from the statistical uncertainties in the images, and are typically smaller than the marker size. The brightness temperature is a factor of 2--3 lower if the angular size of the VLBI observation from \cite{Benz1998} is used, and potentially 100 times lower for an extended magnetosphere.\label{fig:tb}  The dotted line plots the brightness temperature for constant flux density (using the mean value of the flux density), emphasizing that the flux density is similar at all frequencies observed. For the VLA data, the flux density shown is obtained from the peak pixel in the image of the source, while for the ALMA flux density is the peak of a Gaussian fit to the source. The gray scale lines present other observations from the literature for comparison, as described in \ref{fig:flux}}
\end{figure}

Figure \ref{Tb} shows an estimated brightness temperature of all of the bands observed, with flares and ECME bursts removed and plotted as a function of frequency. For reference, the dotted line plots the brightness temperature corresponding to a constant flux density value (chosen as the mean flux density of all the observations), and the gray lines plot the brightness temperature corresponding to the examples from \cite{gudelbenz1996radiospectra} and \cite{Gudel1989} shown in Figure \ref{fig:flux}. The brightness temperatures presented are calculated for a source size equal to the size of the stellar disk. Note that the actual size of the emission region may be much smaller than the stellar disk, if it originates from a small-scale feature, or may be larger than the stellar disk if it originates in an extended large-scale magnetosphere. Thus, the true brightness temperature may be larger or smaller than that presented in Figure \ref{fig:tb}, but the stellar disk size remains an important possible size scale to use for reference here. For this fixed source size, the brightness temperature falls steadily with frequency, as the persistent component of emission is fairly similar from 1--105\,GHz. Although these are not simultaneous measurements, the similarity across all the bands makes it worth discussing the implications for the emission mechanism in a scenario where the persistent component in each observation does reflect a steady flux density across all the days. If these points do approximate the spectral energy distribution of one unifying emission mechanism, the flux density is too constant with frequency to be optically thick.  In \cite{dulk1983AMherculis}, a model of isotropic optically thick gyrosynchrotron emission in a dipolar magnetic field was used to describe the radio emission from the M dwarf component of a cataclysmic variable, where a radius-dependent electron population caused the spectrum to flatten and turnover above 10\,GHz. This model, however falls off too steeply at high and low frequencies and would not produce the flatness observed over the entire 1--105\,GHz range. The flux density is also too flat to be described by optically thin gyrosynchrotron emission from an isotropic distribution of electron pitch angle.  A model analogous to Jupiter's magnetosphere may be required, with an anisotropic electron distribution confined to shells in the magnetosphere \citep[e.g.][]{depater1980-jupiter-thesis}.

\label{Tb}

\section{Conclusions}
In summary, UV Ceti exhibits slowly variable persistent emission at all the frequencies observed as well as flares that suggest a combination of auroral emission processes (ECME) and coronal emission processes (fast-rising exponential decay flares). 
In these observations, temporal variability is detected in every band observed. Fast-rising exponentially-decaying flares are observed at K band, consistent with gyrosynchrotron emission from flares due to small-scale magnetic field activity. Intensely circularly polarized flares are observed at C and S band, consistent with auroral ECME.

The circular polarization fraction appears to vary strongly with rotation phase even at Ka band. Confirming this behavior will require an observation at a wide range of frequencies simultaneously with full rotation-period coverage over multiple rotation periods. The wide frequency coverage required can be achieved using the VLA in sub-array mode, by which different groups of VLA antennas use a different receiver, such that multiple receiver bands can be observed simultaneously.  If the strong variability in the circular polarization fraction at Ka band is due to electron cyclotron maser emission, then the emission occurs in a region where the magnetic field is dominated by small-scale field features that exceed 5\,kG, rather than the dipole component of the magnetic field. Since electron cyclotron maser emission from Jupiter extends to frequencies that require involving higher-order magnetic field terms, this requirement may be reasonable for UV Ceti.  
Finally, the persistent (steady or slowly-varying) emission maintains a fairly similar flux density from 1--105\,GHz, and makes energetic electrons trapped in a radiation belt with an anisotropic pitch-angle distribution an interesting possibility. Finally, the extreme differences between the emission from UV Ceti and its less-active companion emphasize that the phenomena explored in this paper are not ubiquitous across M dwarfs but rather form one possible repertoire of magnetospheric activity.

\vspace{5mm}
\section{Acknowledgements}
KP thanks Ivey Davis and Jackie Villadsen for useful discussions, and Dillon Dong for helpful conversations about VLA data reduction and CASA.
KP was supported by an NSF GRFP fellowship for part of the time that this research was conducted.
This research has made use of data from the OVRO 40-m monitoring program (Richards, J. L. et al. 2011, ApJS, 194, 29), supported by private funding from the California Insitute of Technology and the Max Planck Institute for Radio Astronomy, and by NASA grants NNX08AW31G, NNX11A043G, and NNX14AQ89G and NSF grants AST-0808050 and AST- 1109911.

\bibliography{references}

\end{document}